\documentstyle[12pt,epsf]{article}
 \newcommand {\bi} {\bibitem}
 \newcommand {\be} {\begin{equation}}
\newcommand {\bea} {\begin{eqnarray} \nonumber }
\newcommand {\ee} {\end{equation}}
\newcommand {\eea} {\end{eqnarray}}
 \newcommand {\eps} {\epsilon}
 \newcommand {\si} {\sigma}
\newcommand {\de} {\delta}

\newcommand {\for} {\ \ \ \mbox{for}\ \ }

\newcommand {\cp} {\right)}
\newcommand {\ap} {\left(}

\def \form#1 {eq. (\ref{#1}) }
\def \parziale#1#2  {{\partial {#1} \over \partial {#2}}}

\begin{document}

\title{Non-equilibrium fluctuation dissipation relations in binary
glasses}
\author{
Giorgio
Parisi\\
Dipartimento di Fisica, Universit\`a {\em La  Sapienza},\\ 
INFN Sezione di Roma I \\
Piazzale Aldo Moro, Rome 00185}
\maketitle

\begin{abstract}
We show that at short time aging is realized in a simple model of binary glasses.  We find that below 
the transition point the off-equilibrium correlation functions and response functions are compatible 
with the relations that were originally derived Cugliandolo Kurchan for generalized spin glasses.
\end{abstract}

\section{Introduction}

The behaviour of an Hamiltonian system (with dissipative dynamics) approaching equilibrium is well 
understood in a mean field approach for infinite range disordered systems \cite{CUKU,FM,BCKM}.   In
the low temperature phase the  correlation and response functions satisfy some simple relations
\cite{CUKU}.  We present indications that they are satisfied in the case of binary
glasses. 

Generally speaking in a non equilibrium system it is natural to investigate  the 
properties of the correlation functions and of the response functions.  Let us concentrate our 
attention on a quantity $A(t)$, which depends on the dynamical variables $x(t)$.  Let us suppose 
that the system starts at time $t=0$ from a given initial condition and subsequently it follow the 
laws of the evolution at a given temperature $T$.  If the initial configuration is not at 
equilibrium at the temperature $T$, the system will display an off-equilibrium behaviour.  In many 
case the initial configuration is at equilibrium at a temperature $T'>T$.  In this note we will 
consider only the case $T'=\infty$.

We can define a correlation function
\be
C(t,t_{w}) \equiv <A(t_{w}) A(t+t_{w})>
\ee
and a response function
\be
G(t,t_{w}) \equiv \frac{ \de A(t+t_{w})}{\de \eps(t_{w})}{\Biggr |}_{\eps=0},
\ee
where we are considering the evolution in presence of a time dependent Hamiltonian in which we have
added the term
\be
\int dt \eps(t) A(t).
\ee

The off-equilibrium fluctuation dissipation relation \cite{CUKU} states some properties of the 
correlation functions in the limit $t_{w}$ going to infinity.  The usual equilibrium fluctuation 
dissipation (FDT) relation tell us that
\be G(t)= - \beta \frac{\partial C(t)}{\partial dt}, \ee
where
\be
G(t)=\lim_{t_w \to \infty} G(t,t_w), \ \ C(t)=\lim_{t_w \to \infty} C(t,t_w).
\ee

It is convenient to define the integrated response:
\be
R(t,t_{w})=\int_{0}^{t} d\tau G(\tau,t_{w}),\ \ 
R(t)=\lim_{t_w \to \infty} R(t,t_w),
\ee
i.e.  the response  to a field acting for a time $t$.

 We can also define the quantities
\be
C(\infty) = \lim_{t \to \infty} C(t), \ \ R(\infty) = \lim_{t \to \infty} R(t)=\beta\ap
C(0)-C(\infty)\cp. \ee
The last equation can be naively written also as
\be
R(\infty)=\beta\ap <A^{2}>-<A>^{2} \cp \label{LINEAR},
\ee
Here the brakets denote the usual equilibrium expectation value.

In the study of off-equilibrium spin glasses systems Cugliandolo and Kurchan \cite{CUKU} proposed 
that the response function and the correlation function satisfy the following relations:
\be 
G(t,t_w)\approx-\beta X \ap C(t,t_w) \cp \frac{\partial C(t,t_w)}{\partial t}.
\ee
The previous relation can be also written in the following form
\be
R(t,t_w)\approx \beta \int_{C(t,t_w)}^{C(0,t_{w})}X(C) dC.
\ee
The function $X(C)$ is system dependent and its form tell us many interesting information.  

If $C(\infty)\ne0 $, we must distinguish two regions: (a)
A short time region where $X=1$ (the so called FDT region) and $C>C(\infty)$.
(b) A large time region (usually $t=O(t_w)$ where $C<C(\infty)$ and $X<1$ (the aging region) 
\cite{B,POLI}.

In the simplest non trivial case, i.e.  one step replica symmetry breaking, \cite 
{mpv,parisibook2,kirtir} the function $X(C)$ is piecewise constant, i.e.
\be
X(C)= m \for C<C(\infty),\ \ \ \ \
X(C)= 1 \for C>C(\infty) \label{ONESTEP}.
\ee

In all known cases in which one step replica symmetry holds, the quantity $m$ vanishes linear with 
the temperature at small temperature.  It often happens that $m=1$ at 
$T=T_{c}$.

The previous considerations are quite general.  However the function $X(C)$ is system dependent 
and its form tell us many interesting information.  Systems in which the replica symmetry is not 
broken are characterized by having $m=0$ in the formula \form{ONESTEP} .

Sometime simple aging is also assumed \cite{B}, i.e.  the following the scaling relation 
holds outside the FDT region:
\be
C(t,t_w)=C_{s}\ap{t \over t_{w}}\cp.
\ee

Simple aging may be correct, but it is not a necessary consequence of the previous relations. We
will see that simple aging is satisfied also in this system \cite{PAAGE}).

\section{The model}
The model we consider is 
the following.  We have taken a mixture of soft particles of different sizes.  Half of the particles 
are of type $A$, half of type $B$ and the interaction among the particle is given by the 
Hamiltonian:
\begin{equation}
H=\sum_{{i<k}} \left(\frac{(\si(i)+\si(k)}{|{\bf x}_{i}-{\bf x}_{k}|}\right)^{12},\label{HAMI}
\label{HAMILTONIAN}
\end{equation}
where the radius ($\si$) depends on the type of particles.  This model has been 
carefully studied  in the past \cite{HANSEN3,LAPA}.  It is known that a 
choice of the radius such that $\si_{B}/\si_{A}=1.2$ strongly inhibits crystallisation and the 
systems goes into a glassy phase when it is cooled.  Using the same conventions of the previous 
investigators we consider particles of average diameter $1$, more precisely we set
\begin{equation} 
{\si_{A}^{3}+ 2 (\si_{A}+\si_{B})^{3}+\si_{B}^{3}\over 4}=1.
\label{RAGGI}
\end{equation}
 
Due to the simple scaling behaviour of the potential, the thermodynamic quantities depend only on 
the quantity $T^{4}/ \rho$, $T$ and $\rho$ being respectively the temperature and the density.  For 
definiteness we have taken $\rho=1$.  The model as been widely studied especially for this choice of 
the parameters.  It is usual to introduce the quantity $\Gamma \equiv \beta^{4}$.  The glass 
transition is known to happen around $\Gamma=1.45$ \cite{HANSEN3}.

Our simulation are done using a Monte Carlo algorithm, which is more easy to deal with than 
molecular dynamics, if we change the temperature in an abrupt way.  Each particle is shifted by a 
random amount at each step, and the size of the shift is fixed by the condition that the average 
acceptance rate of the proposal change is about .4.  Particles are placed in  a cubic box with 
periodic boundary conditions.  We start by placing the particles at random and we quench the 
system by putting it at final temperature (i.e.  infinite cooling rate).

The main quantity on which we will concentrate our attention is the asymmetry in the energy (or 
stress):
\be
A=\sum_{{i<k}} \frac{(\si(i)+\si(k))^{12}} {|{\bf x}_{i}-{\bf x}_{k}|^{14}}
\ap 2(x_{i}-x_{k})^{2}-(y_{i}-y_{k})^{2}-(z_{i}-z_{k})^{2}\cp
 =2T_{1,1}-T_{2,2}-T_{3,3}.
\ee
In other words $A$ is a combination of the diagonal components of the stress energy tensor.
If the particles are in a cubic symmetric box, we have that
\be
<A>=0.
\ee
If the box does not have a cubic symmetry the effect of the boundary disappears in the infinite 
volume limit (at fixed shape of the boundary) and we have that the stress density $a$ vanishes in 
this limit:
\be
\lim_{N\to\infty}{<A>\over N} \equiv a =0.
\ee 

 \begin{figure}[htbp]
\epsfxsize=400pt\epsffile[22 206 549 549]{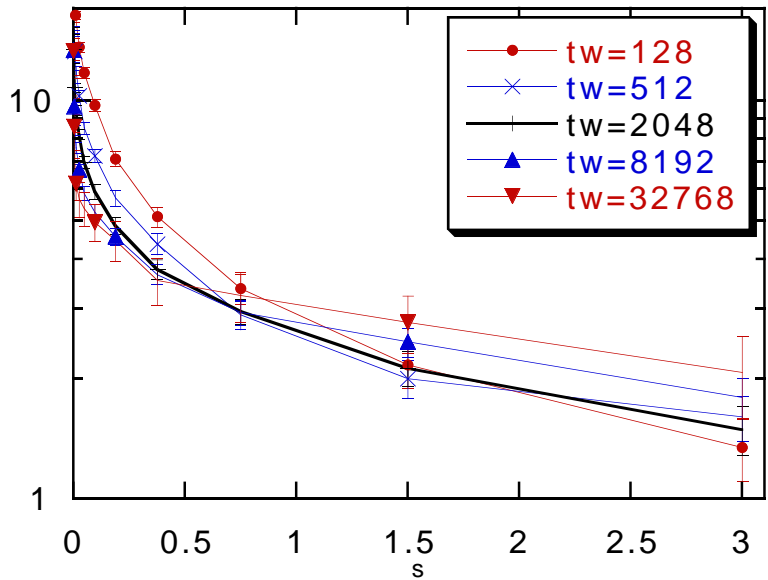}
\caption{On a linear scale we see the response function $S$ at $\Gamma=1.8$ for $N=130$ at
different values of $t_{w}$ (128, 512, 2048, 8912 and 32768) as function of $s \equiv 
t/t_w$.}
\label{RLIN}
\end{figure}

\section{Numerical simulations}

We have done simulations for various values of $N$ ranging from $N=18$ to $N=130$. We report here
only the results for $N=130$ where we have done the average over $250$ samples, those for smaller
systems are qualitatively similar, and can be found in ref \cite{PAAGE}.  The evolution was  done
using the Monte Carlo method, with an acceptance rate fixed around .4.

The limit $t \to \infty $ may be tricky.  All the previous theoretical discussion were done for an 
infinite system.  In practice we need that that $N$ is greater that the time to a given power, the 
exponent of the power being system dependent.  Therefore we cannot strictly take the limit $t
\to \infty $ for finite $N$.  We have to study the behaviour of the system in region of time whose 
size increase with $N$.  Finite size corrections sometimes become much more severe at large times
\cite{ PAAGE}.

We will study the correlation $C(t,t_{w})$ and the response $R(t,t_{w})$.  We introduce the variable 
$s=t/t_{w}$.  According to simple aging the correlation functions should become a function of only 
$s$ in the limit of large times.  Of course the FDT region, which is located at finite $t$ also when 
$t_{w}$ goes to infinity, is squeezed at $s=0$, so that we expect that the function $R$ becomes 
discontinuous at $s=0$.  Moreover the limit $s\to 0$ give us information on the value of $R(\infty)$
\be
\lim_{s \to 0}R(t,t_{w})=R(\infty),
\ee
where it is understood that the limit is done always remaining in the region where $t>>1$.

We have followed a standard procedure \cite{CKR,FRARIE} to measure the off-equilibrium response 
function in simulations: we have kept the system in presence of an external field $\eps$ up to time 
$t_{w}$ and we have removed the field just at this time.

If $\eps$ is sufficient small, we have that the stress as function of time, is related to the 
integrated response $R$ by the relation
\be
{A(t+t_{w}) \over \eps} \equiv S(t+t_{w})=  R(t+t_w,0)-R(t,t_{w})\label{MISURA}
\ee
As far as the limit of $R(t,t_{w})$ when $t\to \infty$ does not depend on $t_{w}$ the quantity 
defined in eq.  (\ref{MISURA}) goes to zero when $t\to \infty$.  In this way we can get the value of 
the integrated response by measuring the stress density as function of time.

\begin{figure}[htbp]
\epsfxsize=400pt\epsffile[22 206 549 549]{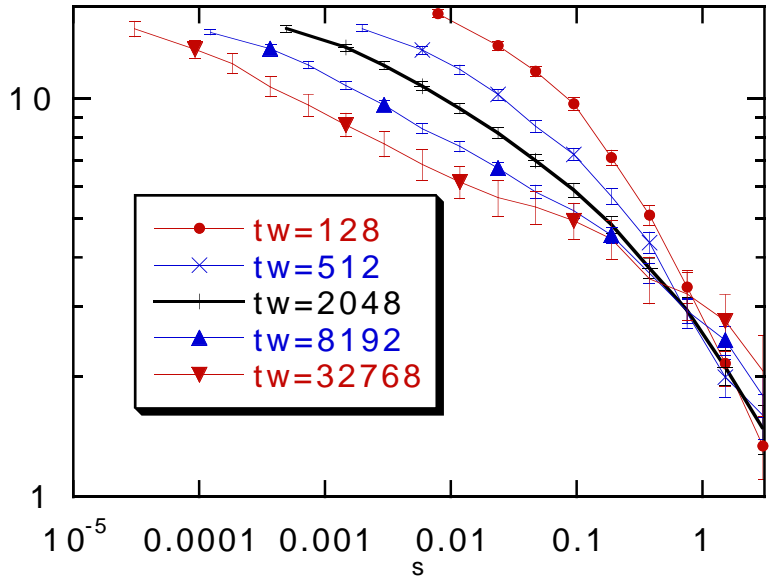}
\caption{On a logarithmic scale we see the response function $S$ at $\Gamma=1.8$ for $N=130$ at
different values of $t_{w}$ (128, 512, 2048, 8912 and 32768) as function of $s \equiv 
t/t_w$.}
\label{RLOG}
\end{figure}
In our case (where we use the stress as a perturbation) the physical interpretation of the procedure 
is quite clear.  We start by putting the systems in a box which is not cubic (because $\eps\ne 0)$, 
but two sides are slightly longer of the third.  At time $t_{w}$ we change the form of the box to a 
cubic one.  In this way we deform the the system and we induce a stress which will be eventually 
decay.  In the high temperature phase, where the system is liquid, the stress will disappear in a 
short time.  On the contrary, in the glassy phase, we shall see that the stress remains for a much 
longer time (as expected
\cite{POLI}) and it shows an interesting aging behaviour.

The choice of $\eps$ is crucial.  In principle its value should be infinitesimal.  However the 
signal is proportional to $\eps$ while the errors are $\eps$ independent.  The errors on the 
response function grow as $\eps^{-1}$.  On the other hand if we take a too large value of $\eps$ we 
enter in a non linear region.   We have taken data at $\eps=.1$ and $\eps=.05$ and we have seen that
there are  some non-linear effects.  No non-linear effects have been detected at $\eps=.02$ and
$\eps=.01$.   All the data we present in this paper come from $\eps=.01$ and they are reasonable free
of systematic  effects.  

We plot the response function $S$ at $\Gamma=1.8$ for $N=130$ at different values of $t_{w}$ (128, 
512, 2048, 8912 and 31768) as function of $s$ on a linear scale (\ref{RLIN}) and on a logarithmic 
scale (\ref{RLOG}).

As we can see the two regions $FDT$ and aging are quite clear.  It is also evident that $S(\infty)$ 
is different from zero at this temperature (it is obviously zero in the high temperature phase.  a 
residual dependence on $t_{w}$.  The short time region is shifted at smaller and smaller $s$ when 
$t_{w}$increases

The crucial step would be now to plot the fluctuations $C$ and the response $S$ together.  The 
previous equation tell us that
\be
{\partial S \over \partial C} = X(C)
\ee
so that it is convenient to plot $S$ versus $C$.  The slope of the function $S(C)$ is thus $X(C)$.

\begin{figure}[htbp]
\epsfxsize=400pt\epsffile[22 206 549 549]{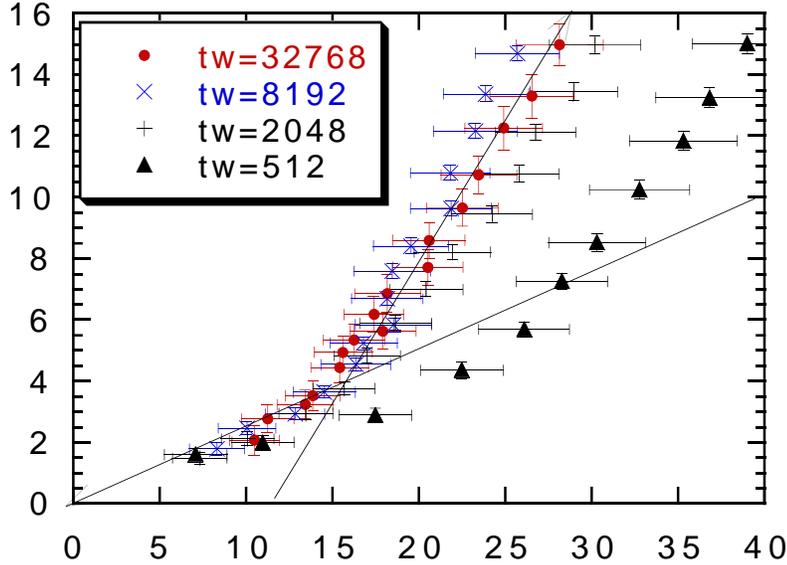}
\caption{The response $S$ as function of $C$ at  $\Gamma=1.8$ for $N=130 $ at
different values of $t_{w}$ (128, 512, 2048, 8912 and 32768).  The two straight lines have slope 1 
and .25.}
\label{CR}
\end{figure}

 In the one step replica symmetry breaking scheme we expect that:
\bea
S= mC \for C< C(\infty),\\
S= C+ (1-m) C(\infty) \for C< C(\infty).
\eea

In fig.  (\ref{CR}) we find the data for the response $S$ as function of $C$ at $\Gamma=1.8$ for 
$N=130 $ at different values of $t_{w}$ ( 512, 2048, 8912 and 32768).  We can see that apart from 
the data at $t_{w}=512$ the data for higher $t_{w}$ are compatible, inside the errors and the 
predicted scaling works quite well.  We can easily see a region where the FDT theorem is valid.  
Beyond this region the data are compatible with a linear dependance of $S$ on $C$ as in the case of 
one step replica symmetry breaking.  The straight line correspond to one step replica symmetry 
breaking with $m=.25$.  This value of $m$ is quite compatible with the value of $m$ ($m=.33\pm.04$) 
found in a different simulation where a different quantity (i.e.  the diffusion) was studied)
\cite{PAAGE}.

The numerical results are in very good agreement with the theoretical expectations and support the 
possibility of one step replica symmetry breaking for real glasses. 

\section* {Acknowledgments} I thank L.
Cugliandolo, S.  Franz, J.  Kurchan and D.  Lancaster for useful discussions.


\begin{thebibliography}{99}
 
\bi{CUKU} L.  F.  Cugliandolo and J.Kurchan, Phys.  Rev.  Lett.  {\bf 71}, 
1 (1993).

\bi{FM} S. Franz and M. M\'ezard {\it On mean-field glassy
dynamics out of equilibrium},  cond-mat 9403004. 

\bi{BCKM} J.-P.  Bouchaud, L.  Cugliandolo, J.  Kurchan, Marc M\'ezard, cond-mat 9511042.

\bibitem{mpv} M.M\'ezard, G.Parisi and M.A.Virasoro, {\sl Spin glass theory and 
beyond}, World Scientific (Singapore 1987).

\bibitem{parisibook2} G.Parisi, {\sl Field Theory, Disorder and
Simulations}, World Scientific, (Singapore 1992).

\bi{kirtir} T. R. Kirkpatrick and D. Thirumalai, Phys.  Rev. {\bf B36}
(1987) 5388 ; T. R. Kirkpatrick and P. G. Wolynes, Phys. Rev. {\bf B36}
(1987) 8552; a review of the results of these authors and
further references can be found in T. R. Kirkpatrick and D. Thirumalai
Transp. Theor. Stat. Phys. {\bf 24} (1995) 927. 


\bi{B} J.-P. Bouchaud; J. Phys. France {\bf 2} 1705, (1992). 

\bi{POLI} L. C. E. Struik; {\it Physical aging in amorphous polymers and other
 materials} (Elsevier, Houston 1978).

\bibitem{PAAGE} G.  Parisi, cond-mat 9701015, 9701100, 9703219.

\bi{HANSEN3}J.-P. Hansen and S. Yip, Trans. Theory and Stat. Phys. {\bf 24}, 1149
(1995).

\bibitem{LAPA} D.  Lancaster and G.  Parisi, cond-mat 9701045.

\bibitem{FRAMAPA} S.  Franz, E.  Marinari, G.  Parisi cond-mat 9506108.

\bi{EA} S. F. Edwards and P. W. Anderson, J. Phys. F {\bf 5}, 965 (1975).

\bi{CKR} L.  F.  Cugliandolo, J.  Kurchan and F.  Ritort; Phys.  Rev.{\bf B49}, 6331 (1994).
     
\bi{FRARIE} S. Franz and H. Rieger Phys.  J. Stat. Phys.  {\bf 79} 749 (1995).

\end{thebibliography}
\end{document}